\documentclass[journal]{IEEEtran}
\usepackage{amsmath}
\usepackage{amsfonts}
\usepackage{amssymb}
\usepackage{amsthm}
\usepackage{graphicx}
\usepackage[caption=false,font=footnotesize]{subfig}
\usepackage{tabularx}
\usepackage{balance}
\usepackage{microtype}
\usepackage[colorlinks,citecolor=blue,linkcolor=blue]{hyperref}
\newtheorem{definition}{Definition}
\begin{document}

\title{\LARGE Electromagnetic Twin: Completing the Wireless World from Sparse Channel Evidence} 
\author{Tuo~Wu,
	Jie Tang,
	Kangda Zhi,
	Junteng Yao,
	Maged Elkashlan,
  	Kin-Fai Tong,~\IEEEmembership{Fellow,~IEEE},\\
  	George K. Karagiannidis,~\IEEEmembership{Fellow,~IEEE},
 and Jinhong Yuan,~\IEEEmembership{Fellow,~IEEE}
\vspace{-9mm}
\thanks{T. Wu and J. Tang are with the School of Electronic and Information Engineering, South China University of Technology, Guangzhou 510640, China (E-mail: $\rm \{wutuo,eejtang\}@scut.edu.cn$). K. Zhi is with the School of Electrical Engineering and Computer Science, Technical University of Berlin, 10623 Berlin (E-mail: $\rm k.zhi@tu$-$\rm berlin.de$). J. Yao is with the Faculty of Electrical Engineering and Computer Science, Ningbo University, Ningbo 315211, China (E-mail: $ \rm  yaojunteng@nbu.edu.cn$). M. Elkashlan is with Queen Mary University of London, United Kingdom (E-mail: $\rm maged.elkashlan@qmul.ac.uk$). K.-F. Tong is with the School of Science and Technology, Hong Kong Metropolitan University, Hong Kong SAR, China (E-mail: $\rm ktong@hkmu.edu.hk$).  	G. K. Karagiannidis is with the Department of Electrical and Computer Engineering, Aristotle University of Thessaloniki, 54124 Thessaloniki, Greece (E-mail: $\rm geokarag@auth.gr$).  J. Yuan is with  the School of Electrical Engineering and Telecommunications, University of New South Wales, Sydney, Australia   (E-mail: $\rm j.yuan@unsw.edu.au$).}
}

\markboth{}{Wu: Electromagnetic Twin from Sparse Channel Evidence}

\maketitle

\begin{abstract}
Acquiring dense channel information over many locations and beams incurs considerable pilot and processing overhead. Radio maps and channel knowledge maps (CKMs) reduce this overhead by reusing site-specific channel information, but their contents must be refreshed when new measurements or environmental observations become available. This paper introduces an \emph{electromagnetic twin} as an updatable digital representation that uses sparse channel evidence to reconstruct the wireless state requested by communication queries. Rather than replacing radio maps or CKMs, the twin uses a CKM as channel memory, combines it with registered scene information, and regenerates its outputs after each evidence update. We instantiate this idea by completing a two-dimensional channel-gain field from sparse samples and an incomplete floor plan. A learned RF completion backbone recovers the main propagation structure, and a lightweight residual adapter tests whether frozen CLIP features provide useful side information. With $4\%$ measured locations and $55\%$ missing semantic objects, the RF backbone attains $4.44$ dB RMSE, compared with $8.29$ dB for CKM interpolation and $8.38$ dB for an incomplete physics prior. Residual adaptation reduces RMSE by a paired mean of $0.135$ dB (95\% confidence interval: $0.100$--$0.169$ dB), but a same-capacity random-feature control is statistically indistinguishable from the CLIP-conditioned adapter. The results therefore support the measurement--update--query loop and lightweight residual correction, while avoiding an unsupported attribution of the correction to visual semantics.
\end{abstract}


\section{Introduction}
\IEEEPARstart{A}{cquiring} channel state information (CSI) over many locations, beams, and frequency resources requires substantial pilot overhead. This is especially costly when the same site is repeatedly queried for coverage, beam selection, outage, or link adaptation. Reusing previously acquired measurements is therefore a central motivation for radio maps and channel knowledge maps (CKMs)~\cite{zengCKM,ckmtutorial}.

A radio map stores a spatial quantity such as received power or path loss, whereas a CKM generalizes this idea to location-indexed gain, delay, angle, path state, best beam, and other reusable descriptors. CKMs have supported hybrid beamforming and channel prediction~\cite{hybridckm,ckmprediction}. Their construction has also progressed from sample-efficient measurement methods~\cite{ckmdata} to convolutional and graph reconstruction~\cite{radiounet,radiogat}, geometry-assisted ray tracing~\cite{indoorrm}, and generative map construction~\cite{genairm}. These methods estimate or store wireless knowledge, but do not by themselves specify when new evidence changes a maintained state and regenerates a requested output. In deployment, stored information should evolve with the physical site. New pilots reduce channel uncertainty; a moved user or object changes blockage and reflection geometry; and a changed carrier, antenna, or beam codebook changes the channel quantity being queried. A digital twin provides the relevant operational principle: physical observations synchronize a maintained digital state, and the synchronized state supports prediction or control~\cite{realtimetwin}. Digital-twin-channel architectures and radio-environment knowledge pools apply this principle to wireless propagation~\cite{dtchannel,rekpool}.

We call the resulting signal-processing object an \emph{electromagnetic twin}: a registered scene/radio state and CKM memory, together with a synchronization model and declared communication queries. A radio map is therefore an output, a CKM supplies persistent location-indexed evidence, and the twin is the update--query process that connects them. It builds on these tools rather than renaming or replacing them. The title phrase \emph{wireless world} is deliberately query-bounded: it denotes the channel quantities needed by the supported service, not every electromagnetic variable. Definition~1 formalizes this distinction.

\begin{figure*}[t]
\centering
\includegraphics[width=.70\textwidth]{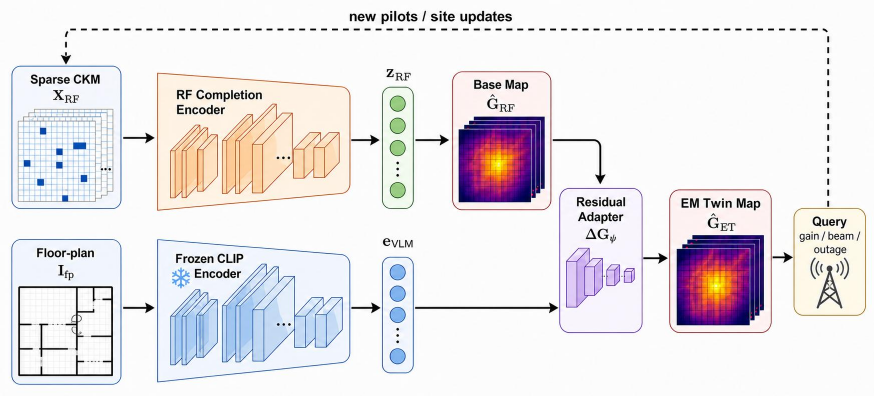}
\caption{Electromagnetic-twin evidence and update flow. The left panels show sparse sounded points and the registered floor plan; the middle implements the fixed learned update; the right returns a gain-map query. The dashed path denotes later pilot or site evidence.}
\label{fig:framework}
\end{figure*}

This paper instantiates the idea through one reproducible update--query cycle. Sparse gain pilots update CKM memory, an incomplete floor plan provides registered scene information, and the query returns a complete fixed-height gain map. The main reconstruction is produced by a convolutional--Transformer RF backbone; a frozen vision--language model (VLM) conditions a small residual adapter, and a same-capacity random-feature control tests whether any improvement is specifically attributable to visual semantics. Fig.~\ref{fig:framework} previews the observed evidence, learned update, and returned map. The contribution is thus not another stand-alone map estimator. It is a minimal evidence--completion--query--feedback interface with explicit scene, memory, update, and query variables; a concrete fusion model for sparse CKM samples, incomplete geometry, and a propagation prior; and a paired evaluation under nested pilots and missing objects that separates residual capacity from semantic conditioning.

\section{System Model and Electromagnetic-Twin Update}
\subsection{Wireless Environment and CKM Measurements}
Consider a fixed-height receiver plane over an indoor wireless service area $\Omega\subset\mathbb R^2$. It is discretized as $\mathcal P=\{\mathbf p_{h,w}:1\leq h\leq H,\,1\leq w\leq W\}$, where $H$ and $W$ are the grid height and width. The registered scene/radio state $\mathcal E_t$ contains the floor layout, object and material labels, base-station (BS) location, carrier and antenna configuration, and receiver height at update index $t$. Let $g_t(\mathbf p)$ denote the channel gain from the BS to $\mathbf p\in\mathcal P$. The left side of Fig.~\ref{fig:framework} illustrates the sparse sounded points and incomplete scene available to the receiver.

Only locations in $\mathcal S_t\subset\mathcal P$ are sounded at update $t$. A measured gain is
\begin{align}
\tilde g_t(\mathbf p)=g_t(\mathbf p)+n_t(\mathbf p),\quad \mathbf p\in\mathcal S_t,
\end{align}
where $n_t(\mathbf p)$ is measurement noise. The CKM memory is
\begin{align}
\mathcal M_t=\{(i,\mathbf p,\tilde g_i(\mathbf p)): \mathbf p\in\mathcal S_i,\ i\le t\},
\end{align}
which retains the update index, location, and measured gain from successive soundings. This is the gain slice of a general CKM, which may instead store CSI, delay, angle, path state, or best beam. We study gain because the declared query is the matrix $\mathbf G_t\in\mathbb R^{H\times W}$ with $[\mathbf G_t]_{h,w}=g_t(\mathbf p_{h,w})$. For a best-beam query, each scalar gain record would be replaced by a beam-indexed gain vector and the query would return its maximizing beam; that extension is not evaluated here.

\subsection{From CKM Memory to an Updatable Twin}
A CKM stores and retrieves site-specific channel knowledge. An electromagnetic twin adds an explicit update step: when new pilots or scene observations arrive, the stored evidence is incorporated and the requested wireless output is regenerated.

\begin{samepage}
\begin{definition}[Minimum electromagnetic twin]
At update index $t$, an electromagnetic twin is
\begin{align}
\mathcal T_t=\{\mathcal E_t,\mathcal M_t,\mathcal F_\theta,\mathcal Q\},
\end{align}
where $\mathcal E_t$ is the registered scene/radio state, $\mathcal M_t$ is CKM memory, $\mathcal F_\theta$ is the reconstruction model with trained parameters $\theta$, and $\mathcal Q$ is the set of supported query operators. Here, model weights are fixed during online updates and $\mathcal Q=\{\mathcal Q_g\}$ contains only the gain-map query.
\end{definition}
\end{samepage}

Let $\Delta\mathcal E_t$ contain newly registered or changed object, material, or radio-configuration entries, and let $\Delta\mathcal M_t$ contain the new indexed pilot records. The synchronization operator $\mathcal U_\theta$ merges these increments with $\mathcal E_{t-1}$ and $\mathcal M_{t-1}$ and reruns $\mathcal F_\theta$; it does not retrain $\theta$. The state update and gain-map query are
\begin{align}
\mathcal T_t&=\mathcal U_\theta(\mathcal T_{t-1};\Delta\mathcal E_t,\Delta\mathcal M_t),\\
\hat{\mathbf G}_t&=\mathcal Q_g(\mathcal T_t).
\end{align}
In plain terms, new evidence changes the registered state or CKM memory, the completion model is evaluated again, and the refreshed gain map becomes available to wireless applications. This update--query operation, rather than the map alone, is what makes the maintained object a twin.

\begin{table}[!b]
\centering
\scriptsize
\caption{Roles of radio map, CKM, and electromagnetic twin.}
\label{tab:comparison}
\begin{tabularx}{\columnwidth}{lXX}
\hline
Object & Primary representation & Defining operation\\
\hline
Radio map & Spatial field of RSS, gain, path loss, or spectrum occupancy & Estimate or query one wireless field\\
CKM & Location-indexed channel descriptors and decisions & Store, infer, and retrieve channel knowledge\\
EM twin & Registered scene/radio state, CKM memory, update model, and query interface & Update stored state and regenerate wireless outputs\\
\hline
\end{tabularx}
\end{table}

Table~\ref{tab:comparison} summarizes the operational distinction: a radio map is a spatial output, a CKM is reusable channel memory, and an electromagnetic twin maintains those objects through explicit synchronization and query operations.

\subsection{Gain-Map Update Studied in This Paper}
For the considered update, the observed scene is a four-channel raster $\mathbf S_{{\rm inc},t}\in\mathbb R^{H\times W\times4}$ containing walls, reflectors, the BS, and nominal wall loss. The binary pilot mask $\mathbf M_t$, normalized measured-gain raster $\mathbf Y_t$, and normalized physics map $\mathbf G_{{\rm phy},t}$ belong to $\mathbb R^{H\times W}$. Unmeasured entries of $\mathbf Y_t$ are zero but are distinguished from measured low gain by $\mathbf M_t$. Concatenating these seven channels gives
\begin{align}
\mathbf X_t&=[\mathbf S_{{\rm inc},t},\mathbf M_t,\mathbf Y_t,\mathbf G_{{\rm phy},t}],\\
\hat{\mathbf G}_t&=f_\theta(\mathbf X_t),
\end{align}
the model reconstructs all $H\times W$ normalized gains, which are converted back to dB for root-mean-square error (RMSE) evaluation. When pilots arrive, the corresponding mask and gain entries are refreshed and the same fixed model regenerates $\hat{\mathbf G}_t$. This is a minimal, replayable synchronization event rather than online weight adaptation or a claim of real-time deployment.

\section{Proposed Electromagnetic-Twin Map Update}
The proposed update model uses three sources. Measured CKM samples tie the result to the current site, a simple propagation model supplies physical structure, and learned RF/visual features estimate the remaining spatial field.

\subsection{Measurement and Physics Representations}
For a data-only reference, let $\mathcal S_t^{\rm mem}$ be the distinct locations represented in $\mathcal M_t$ and let $\bar g_t(\mathbf s)$ be the latest stored gain at $\mathbf s$. Assign $w_{\mathbf s}(\mathbf p)=(\|\mathbf p-\mathbf s\|_2+\epsilon)^{-1}$ so that nearby records receive larger weight. Inverse-distance CKM interpolation gives
\begin{align}
\hat g_{{\rm CKM},t}(\mathbf p)=
\frac{\sum_{\mathbf s\in\mathcal S_t^{\rm mem}}w_{\mathbf s}(\mathbf p)\bar g_t(\mathbf s)}
{\sum_{\mathbf s\in\mathcal S_t^{\rm mem}}w_{\mathbf s}(\mathbf p)}.
\end{align}
Thus, $\hat g_{{\rm CKM},t}$ is the measurement-only gain estimate; $\epsilon=0.25$ grid cell prevents a singularity. Independently, the incomplete floor plan produces a rough physics map. Let $\bar g_{0,t}(\mathbf p)$ be the direct-path gain in dB and $\bar g_{r,t}(\mathbf p)$ the gain of a candidate single-reflection path $r\in\mathcal R_t(\mathbf p)$:
\begin{align}
\bar g_{0,t}(\mathbf p)
&=g_{\rm ref}-10\gamma\log_{10}d_0(\mathbf p)-L_{w,0}(\mathbf p),\\
\bar g_{r,t}(\mathbf p)
&=g_{\rm ref}-10\gamma\log_{10}d_r(\mathbf p)-L_{\rm ex}\nonumber\\
&\quad-\kappa\!\left[L_{w,r}^{(1)}(\mathbf p)+L_{w,r}^{(2)}(\mathbf p)\right]+G_r,\\
g_{{\rm phy},t}(\mathbf p)
&=10\log_{10}\!\left(10^{\bar g_{0,t}(\mathbf p)/10}
+\sum_{r\in\mathcal R_t(\mathbf p)}10^{\bar g_{r,t}(\mathbf p)/10}\right).
\end{align}
Here $g_{\rm ref}$ and $\gamma$ are the 1-m reference gain and path-loss exponent. The path lengths $d_0$ and $d_r$, reflector set $\mathcal R_t(\mathbf p)$, and crossed-wall losses are computed geometrically from the registered raster: $L_{w,0}$ is the direct-path wall loss, while $L_{w,r}^{(1)}$ and $L_{w,r}^{(2)}$ belong to the two reflected segments. The material values, reflector gain $G_r$, and excess loss $L_{\rm ex}=6$ dB are the calibrated simulation parameters specified in Section~IV-A. The effective factor $\kappa=0.45$ accounts for oblique reflected-segment interaction under the coarse raster/material model; it is not a universal material constant. Direct and reflected powers are summed in the linear domain. The normalized $g_{{\rm phy},t}$ forms $\mathbf G_{{\rm phy},t}$ and becomes inaccurate when registered objects or material labels are missing.

\subsection{Learned RF Completion Backbone}
The RF backbone receives the previously defined input
\begin{align}
\mathbf X_t=[\mathbf S_{{\rm inc},t},\mathbf M_t,\mathbf Y_t,\mathbf G_{{\rm phy},t}],
\end{align}
The seven channels have already been defined in Section~II-C. The mask lets the network distinguish an unmeasured zero from a measured low-gain value. The encoder uses width 72 and two stride-2 downsampling stages; four six-head Transformer layers process the quarter-resolution spatial tokens, and two transposed-convolution stages restore the grid. A one-channel head predicts the base map
\begin{align}
\hat{\mathbf G}_{{\rm RF},t}=f_\Theta(\mathbf X_t).
\end{align}
The backbone is trained over randomized sites using
\begin{align}
\mathcal L_{\rm RF}(\Theta)=
\|\hat{\mathbf G}_{\rm RF}-\mathbf G\|_1
+\mu\|\hat{\mathbf G}_{\rm RF}-\mathbf G\|_2^2,\quad \mu=0.5.
\end{align}
The $\ell_1$ term is robust to localized shadow-boundary errors, whereas the squared $\ell_2$ term penalizes distributed bias and stabilizes optimization; $\mu=0.5$ fixes their relative scale. Training over randomized indoor sites makes $f_\Theta$ reusable across the tested site distribution rather than fitted to one map. We call it an RF backbone, not a foundation model, because broad cross-domain pretraining is not evaluated.

\subsection{Frozen CLIP Residual Adapter}
The incomplete scene and pilot positions are rendered as a $224\times224$ floor-plan image $\mathbf I_{{\rm fp},t}$: gray intensity carries wall loss, orange marks reflectors, red marks the BS, and blue marks sounded locations. A frozen CLIP encoder~\cite{clip} extracts
\begin{align}
\mathbf e_{{\rm clip},t}=\mathrm{CLIP}_{\rm img}(\mathbf I_{{\rm fp},t}).
\end{align}
The final prediction is
\begin{align}
\hat{\mathbf G}_{{\rm ET},t}=\hat{\mathbf G}_{{\rm RF},t}
+\alpha\tanh r_\psi(\mathbf X_t,\hat{\mathbf G}_{{\rm RF},t},\mathbf e_{{\rm clip},t}),
\quad \alpha=0.25,
\end{align}
The 512-dimensional CLIP vector is projected to eight channels and broadcast over the grid. The adapter $r_\psi$ concatenates those channels with $\mathbf X_t$ and $\hat{\mathbf G}_{{\rm RF},t}$, applies two 48-channel convolutions, and outputs a one-channel correction. The fixed $\alpha=0.25$ bounds this normalized residual. During adapter training, $f_\Theta$ and CLIP are frozen and only $r_\psi$ is optimized. CLIP therefore conditions, rather than solves, propagation; a same-capacity random-feature adapter tests whether conditioning contributes beyond residual capacity.

Fig.~\ref{fig:framework} and Algorithm~1 separate offline learning from online synchronization. Offline, the backbone and then the residual adapter are trained. Online, their weights remain fixed: new pilots refresh CKM evidence, the physics raster is recomputed if the scene changes, and one forward pass regenerates $\hat{\mathbf G}_{{\rm ET},t}$ for the gain-map query.

\begin{center}
\scriptsize
\begin{tabularx}{\columnwidth}{lX}
\hline
\multicolumn{2}{c}{\textsc{Algorithm 1: EM-Twin Training and Online Update}}\\
\hline
\textbf{Input} & Training scenes; current CKM samples $(\mathbf M_t,\mathbf Y_t)$; incomplete scene $\mathbf S_{{\rm inc},t}$.\\
\textbf{Output} & Updated twin gain map $\hat{\mathbf G}_{{\rm ET},t}$.\\
\hline
1 & \textbf{Offline:} train RF backbone $f_\Theta$ on randomized scenes using $\mathcal L_{\rm RF}$.\\
2 & Freeze $f_\Theta$/CLIP and train residual adapter $r_\psi$ on target maps.\\
3 & \textbf{Online:} append new pilots to CKM memory $(\mathbf M_t,\mathbf Y_t)$.\\
4 & Compute $\mathbf G_{{\rm phy},t}$ and build $\mathbf X_t=[\mathbf S_{{\rm inc},t},\mathbf M_t,\mathbf Y_t,\mathbf G_{{\rm phy},t}]$.\\
5 & Obtain base map $\hat{\mathbf G}_{{\rm RF},t}=f_\Theta(\mathbf X_t)$ and frozen CLIP feature $\mathbf e_{{\rm clip},t}$.\\
6 & Predict residual $\Delta\mathbf G_t=r_\psi(\mathbf X_t,\hat{\mathbf G}_{{\rm RF},t},\mathbf e_{{\rm clip},t})$.\\
7 & Update $\hat{\mathbf G}_{{\rm ET},t}=\hat{\mathbf G}_{{\rm RF},t}+0.25\tanh(\Delta\mathbf G_t)$ and answer queries.\\
\hline
\end{tabularx}
\end{center}

\section{Simulation and Discussion}
\begin{figure*}[t]
\centering
\subfloat[Ground truth.\label{fig:surface_gt}]{%
\includegraphics[width=0.235\textwidth]{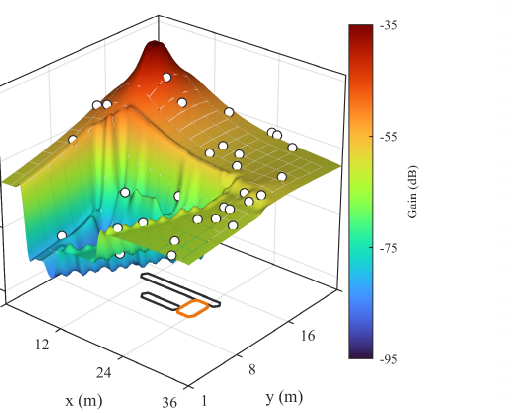}}
\hfill
\subfloat[Map-only CKM (9.49 dB).\label{fig:surface_ckm}]{%
\includegraphics[width=0.235\textwidth]{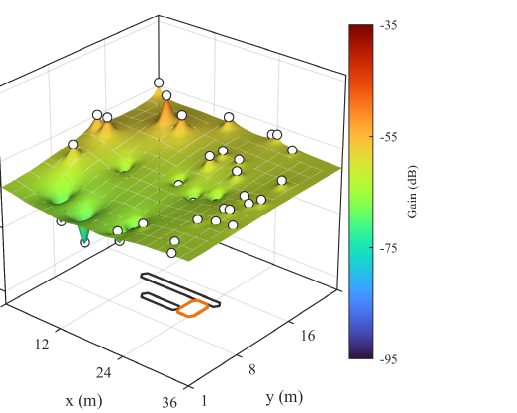}}
\hfill
\subfloat[Physics prior (9.71 dB).\label{fig:surface_phy}]{%
\includegraphics[width=0.235\textwidth]{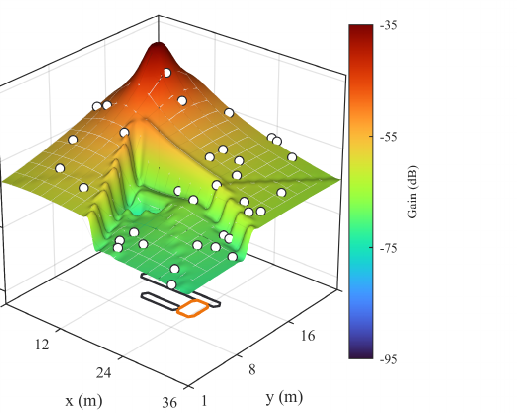}}
\\[-0.5ex]
\subfloat[RF backbone (5.19 dB).\label{fig:surface_fm}]{%
\includegraphics[width=0.235\textwidth]{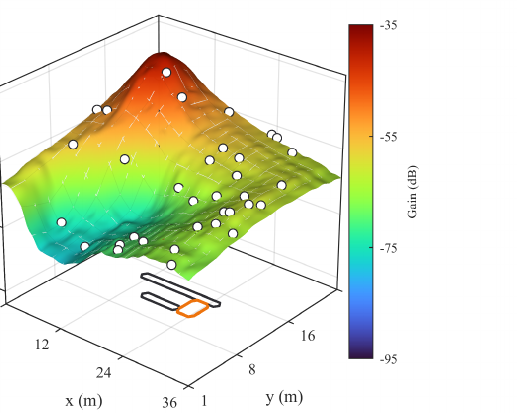}}
\hspace{0.035\textwidth}
\subfloat[CLIP-conditioned ET (5.06 dB).\label{fig:surface_clip}]{%
\includegraphics[width=0.235\textwidth]{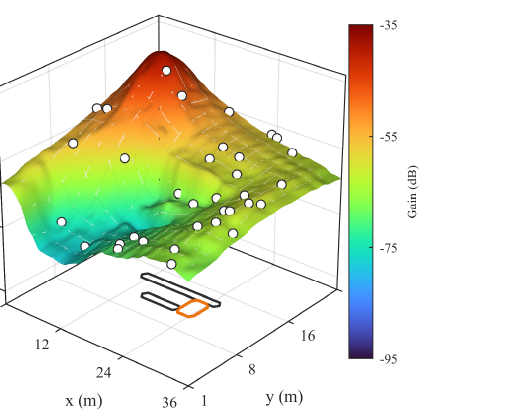}}
\caption{Representative fixed-height gain-map completion results. White circles mark pilots, black/orange contours mark walls/reflectors, and the red star marks the BS; parentheses report single-scene RMSE where applicable, and surface elevation denotes channel gain.}
\label{fig:surface_comparison}
\end{figure*}

\subsection{Setup}
We simulate a $36\,\mathrm{m}\times24\,\mathrm{m}$ indoor area on a fixed-height 1-m grid. Each scene contains one BS, $3$--$6$ walls, $1$--$3$ reflectors, and 12 beam sectors. The channel model uses a $-34$ dB reference gain at 1 m and $21\log_{10}d$ dB distance loss. Each crossed wall adds $10$--$23$ dB loss. A reflected path includes 6-dB excess loss, reflector gain, and $\kappa$-scaled wall losses on both segments, with the same $\kappa=0.45$ used in the simulator and physics prior; LoS and reflected powers are summed in the linear domain. CKM measurements contain independent 0.8-dB Gaussian noise, and the observed floor plan is degraded by removing semantic objects.

The $3{,}130{,}993$-parameter RF backbone is trained for 420 steps with batches of 18, corresponding to 7,560 procedurally generated training scenes. The CLIP-conditioned and random-feature adapters each contain $95{,}513$ trainable parameters and use 420 steps with batches of 14, or 5,880 generated scenes. All networks use AdamW with learning rate $2\times10^{-4}$, weight decay $10^{-4}$, the loss weight $\mu=0.5$, and mixed-precision CUDA. The \texttt{openai/clip-vit-base-patch32} encoder and RF backbone are frozen during adapter training; training the CLIP adapter requires 68.54 s on an NVIDIA GeForce RTX 5080.

Training, validation, and test scenes are generated from disjoint pseudorandom streams. The training realizations are not reused in the 192-scene validation set or the 192-scene test set, and the validation and test seed ranges are also disjoint. The number of optimization steps is fixed before evaluation; validation checks convergence and is not used for post-hoc checkpoint selection. All reported curves average the same 192 test scenes, and all method comparisons within a point use identical scenes, missing-object realizations, pilot locations, and measurement noise.

Two tests are performed. The \emph{pilot-update test} fixes the missing-object ratio at $55\%$ and appends nested pilot sets from $1\%$ to $8\%$ of grid points. It tests whether the twin improves when CKM memory receives new measurements. This is iterative evidence assimilation in a static scene, not online weight adaptation or time-varying environment tracking. The \emph{scene-degradation test} fixes pilots at $4\%$ and removes $0\%$--$75\%$ of semantic objects. It tests whether measured channels can stabilize the update when the floor plan is incomplete.

The compared methods are:
\begin{itemize}
\item \textbf{Map-only CKM:} measurement-only interpolation, without scene information.
\item \textbf{Incomplete physics prior:} scene-only propagation calculation, without learned completion.
\item \textbf{RF backbone:} learned fusion of pilots, scene raster, and physics map, without a residual adapter.
\item \textbf{Random-feature adapter:} the same residual architecture as the CLIP variant, conditioned on independent normalized Gaussian features.
\item \textbf{CLIP-conditioned ET:} the complete update with frozen CLIP image features and residual correction.
\end{itemize}

\begin{figure}[t]
\centering
\includegraphics[width=0.58\columnwidth]{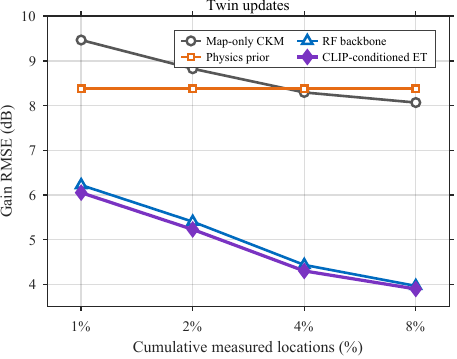}
\caption{Gain RMSE versus cumulative measurements with $55\%$ missing scene objects.}
\label{fig:sparse_curve}
\end{figure}

\begin{figure}[t]
\centering
\includegraphics[width=0.58\columnwidth]{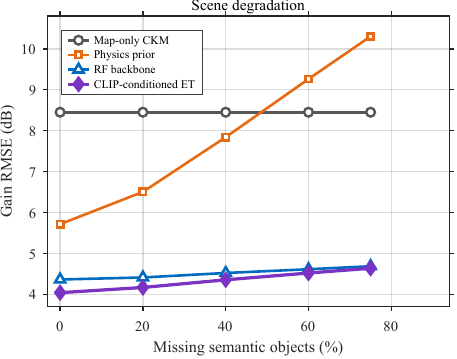}
\caption{Gain RMSE versus missing scene objects with $4\%$ measured locations.}
\label{fig:missing_curve}
\end{figure}

\subsection{Main Results}
Table~\ref{tab:mainresult} reports the common operating point of $4\%$ pilots and $55\%$ missing objects. The data-only and scene-only baselines both exceed 8 dB RMSE. Fusing the two sources with the RF backbone reduces RMSE to 4.44 dB, showing that the main gain comes from learned RF completion. Both residual variants reach 4.30 dB.

\begin{table}[t]
\centering
\caption{RMSE at $4\%$ measured locations and $55\%$ missing objects; CLIP versus random features: $p=0.993$.}
\label{tab:mainresult}
\scriptsize
\begin{tabularx}{\columnwidth}{Xcc}
\hline
Method & RMSE (dB) & Std. (dB)\\
\hline
Map-only CKM & 8.29 & 2.36\\
Incomplete physics prior & 8.38 & 4.51\\
RF backbone & 4.44 & 1.63\\
Random-feature adapter & \textbf{4.30} & 1.58\\
CLIP-conditioned ET & \textbf{4.30} & \textbf{1.55}\\
\hline
\end{tabularx}
\end{table}

Because every method is evaluated on the same scene realizations, the adapter contribution is tested on per-scene RMSE differences rather than by comparing two independent standard errors. Relative to the RF backbone, the CLIP-conditioned adapter improves RMSE by $0.135$ dB on average, with a 95\% Student-$t$ confidence interval of $[0.100,0.169]$ dB and paired $p=8.5\times10^{-13}$. However, the CLIP-conditioned and random-feature adapters differ by less than $0.001$ dB on average, with 95\% confidence interval $[-0.013,0.013]$ dB and paired $p=0.993$. The statistically supported statement is therefore that the residual adapter improves the frozen RF backbone; this experiment does not isolate an additional benefit from CLIP semantics.

Fig.~\ref{fig:surface_comparison} explains these RMSE trends spatially. CKM interpolation is smooth but misses sharp shadow regions; the physics prior creates geometry-dependent structure but uses incomplete objects. The RF backbone recovers the large-scale gain topology, and residual correction refines it. All panels share one 2D receiver plane; vertical relief denotes gain.\\

Fig.~\ref{fig:sparse_curve} directly evaluates four update rounds. Each round retains all earlier pilots and appends new measurements. From $1\%$ to $8\%$ cumulative pilots, the CLIP-conditioned ET improves monotonically from 6.05 dB to 3.90 dB; CKM interpolation improves from 9.47 dB to 8.07 dB. The model weights are unchanged, so the gain comes from updating CKM evidence rather than retraining. The random-feature curve is omitted for clarity because it nearly coincides with the CLIP-conditioned curve.

Fig.~\ref{fig:missing_curve} isolates scene uncertainty. The physics prior degrades from 5.72 dB with complete geometry to 10.31 dB at $75\%$ missing objects. The learned methods remain near 4--5 dB because pilots still describe the current radio site. The residual variant stays close to the RF backbone across the sweep; together with the random-feature control, this behavior supports residual correction but not a separate semantic-gain claim.

\subsection{Interpretation of the RF and Residual Components}
The results support two conclusions. First, CKM pilots carry current site information and the RF backbone performs most of the spatial reconstruction. Second, a small residual stage can correct the frozen backbone, but the present CLIP representation is not demonstrably better than a random conditioning vector. This distinction matters: a pretrained visual encoder may still help under richer images, cross-layout transfer, or textual material descriptions, but such a semantic claim requires a control-separated gain. In the current study, the electromagnetic-twin update is driven primarily by wireless measurements, scene channels, and learned RF structure.

\section{Conclusion}
This paper presented electromagnetic twin as an updatable digital representation of site-specific wireless state. CKM stores successive channel measurements, registered scene information describes the physical site, and an update model regenerates radio maps or other communication outputs when new evidence arrives. We instantiated this measurement--update--query process through fixed-height gain-map reconstruction. Sparse pilots anchored the result to the current site, and the RF completion backbone supplied most of the accuracy improvement over measurement-only and scene-only baselines. A lightweight residual stage produced a statistically significant paired improvement over the frozen backbone; however, a same-capacity random-feature control showed that this gain cannot yet be attributed to CLIP semantics. The nested-pilot experiment further showed improvement without model retraining as CKM evidence accumulated, while the scene-degradation test demonstrated robustness to missing objects. These results turn electromagnetic twin from a general digital-twin idea into a concrete, reproducible, and appropriately bounded wireless-communication workflow.

\balance

\end{document}